\documentclass[twocolumn,aps,prl,showpacs]{revtex4-1}
\usepackage{graphicx,bm,amssymb,amsmath}
\usepackage{color}

\usepackage{array}   

\begin{document}
\title{Stimulated Binding of Polymer Chains in Narrow Tube Confinement}

\author{Gillian M. Fraser$^1$, M\'arcio Sampaio Gomes Filho$^{2,3}$, and Eugene M. Terentjev$^2$}

\affiliation{$^1$Department of Pathology, University of Cambridge, Tennis Court Road, Cambridge CB2 1QP, UK}
\affiliation{$^2$Cavendish Laboratory, University of Cambridge, J.J. Thomson Avenue, Cambridge CB3 0HE, UK}
\affiliation{$^3$Instituto de F\'isica, Universidade de Bras\'ilia, Bras\'ilia-DF, Brazil}

\begin{abstract}
\noindent  In biology, there are several processes in which unfolded protein chains are transported along narrow-tube channels. Normally, unfolded polypeptides would not bind to each other. However, when the chain entropy is severely reduced in the narrow channel, we find that polymer chains have a propensity to bind, even if there is no great potential energy gain in doing so. We find the average length of binding $m^*$ (the number of monomers on overlapping chain ends that form bonds) and the critical tube diameter at which such constrained binding occurs. Brownian dynamics simulations of tightly confined chains, confirm the theoretical arguments and demonstrate chain binding over the characteristic length  $m^*$, changing in tubes of different diameter.
\end{abstract}

\pacs{36.20.Ey, 87.15.hp}
\maketitle
\flushbottom

Much of the physics of polymer chains is determined by entropic effects; entropic barriers associated with chain configurations are widely recognized to control polymer behaviour in narrow pores \cite{Muthu1989,Park1998,DiMarzio1997,Muthu2003}. Here we consider a related problem of how confinement in a narrow tube could precipitate binding of polymer chains, even though in free unconstrained conditions they would not have a propensity to bind. The problem is motivated by a set of remarkable biological structures that facilitate movement of nucleic acids or polypeptides through narrow channels: (i) contractile tail bacteriophages inject their nucleic acid genome into host bacterial cells through their hollow tail tube \cite{phage}; (ii) bacterial type IV secretion systems transfer nucleic acids between bacteria and proteins into eukaryotic host cells \cite{Galan2018}; (iii) ribosomes release growing polypeptides through the ribosomal exit tunnel where, for some proteins, the beginning of folding is registered while in confinement \cite{ribosome1,ribosome2}. Finally, bacterial type III secretion systems export unfolded polypeptides that assemble into cell-surface nanomachines that inject bacterial proteins into eukaryotic cells \cite{Galan2018} or, in the case of bacterial flagella, facilitate cell motility \cite{Fraser2014}. During flagella biogenesis, individual subunits synthesised inside the cell need to be unfolded and threaded through a narrow channel that runs the full length of the growing structure, to be assembled at the distal tip \cite{Macnab2003,Berg2012,Fraser2014}.  

For flagella assembly, two mechanisms of subunit transport along the narrow channel have been investigated. One relies on single-file diffusion \cite{Berg2013}, which requires `pushing' a column of subunits through the narrow channel, against tremendous resistance once the crowded regime sets in \cite{Lappala}. The other mechanism suggests that unfolded flagellar subunits form a linked chain along the whole length of the flagellar channel, which is then `pulled' to the site of subunit assembly at the flagellum tip \cite{Evans2013}. Here, we propose a mechanism by which unfolded polymers would strongly bind to form ‘head-to-tail’ links, when constrained in a channel. 

\begin{figure}
\includegraphics[width=0.8\columnwidth]{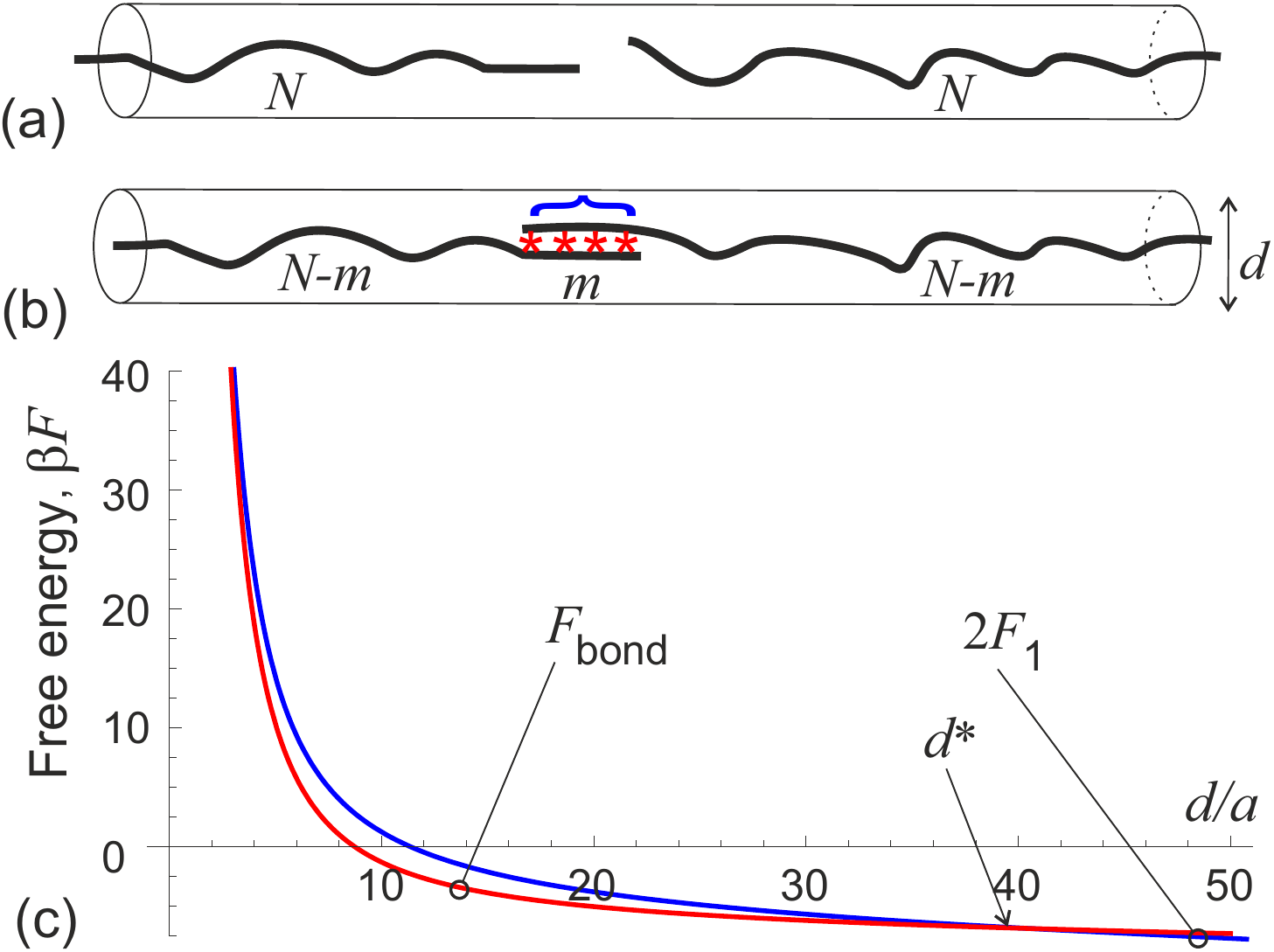}
\caption{\label{fig1}  (a) Two polymer chains of contour length $Na$ confined in a narrow tube of diameter $d$. (b) The same two chains bonded over $m$ units. (c) Comparison of equilibrium free energies $2F_1$ and $F_\mathrm{bond}$, corresponding to the sketches in (a) and (b), for the bonding strength $\beta \Delta=1$ and $m=5$. At large $d$ the chains prefer to be separate, while in a tight tube $d<d^*$ the bonded configuration has lower free energy.}
\end{figure}

The properties of polymer chain ($N$ segments of size $a$) constrained in a narrow channel are very well known, starting from the seminal work of Casassa \cite{Casassa1967} and Edwards \cite{Edwards1969,DoiEdwards}, and extensively reviewed in many subsequent publications \cite{DeGennes,Muthu2003,Rafael2006}. The free energy excess contains two contributions: the `ideal gas' motion of the chain centre of mass along the tube, and the reduction in configurational entropy under constraint:
\begin{equation}
\frac{F_1}{k_BT}\sim -\ln \left(\frac{d}{a} \right)^2 + \left( \frac{R_g}{d} \right)^2 =-2 \ln \left(\frac{d}{a} \right) + \frac{\pi^2 N}{3} \left( \frac{a}{d} \right)^2,
\end{equation}
where the second expression is for the ideal Gaussian chain, with the radius of gyration $R_g=N^{1/2}a$. This is good enough for our purposes; in fact, the commonly used model of self-avoiding chain in good solvent might be less appropriate for a case of protein unfolded in a channel \cite{Evans2013}. When two such chains are in the channel, Fig. \ref{fig1}(a), the free energy is additive: $2F_1$. When these two chains are instead bonded over the length of $m$ overlapping units, see Fig. \ref{fig1}(b), the corresponding free energy has only one centre-of-mass term. The configurational free energy has $2(N-m)$ monomers constrained in a tube of diameter $d$, and strictly speaking, $m$ monomers constrained in a smaller diameter $(d-a)$:
\begin{eqnarray}
\frac{F_\mathrm{bond}}{k_BT} &=&  -\frac{m\Delta}{k_BT}- 2 \ln \left(\frac{d}{a} \right)  \\
&& +\frac{2\pi^2}{3}(N-m) \left( \frac{a}{d} \right)^2+\frac{\pi^2m}{3} \left( \frac{a}{d-a} \right)^2, \nonumber
\end{eqnarray}
where $\Delta$ is the potential energy gain on making one bond (see \cite{binding} and many subsequent studies of this energy for aminoacid residues in contact). We shall be interested in the situation when this gain is small (or non-existent), so that the monomers would not bond in the free-chain conditions. Figure \ref{fig1}(c) shows these two free energies compared, as a function of changing tube diameter. What we discover, is that at large $d$ (when the entropic constraint is weak), the free energy of two independent chains is lower, but as $d$ becomes smaller -- the free energy of the bonded pair is lower. In fact, the free energy difference $\Delta F=F_\mathrm{bond}-2F_1$ no longer depends on the overall chain length $N$, and reveals the key equilibrium effect:
\begin{equation}
\frac{\Delta F}{k_BT}=  2 \ln \left(\frac{d}{a} \right) - m \left[ \frac{\pi^2a^2(d^2 - 4ad +2a^2)}{3d^2(d-a)^2}+ \beta \Delta  \right] , 
\label{DF}
\end{equation}
with $\beta=1/k_BT$. We see that the ideal-gas entropy `wins' at large $d$, but at $(d-a)\rightarrow 0$ the confinement effect enforces the binding via the negative-$m$ contribution. Even when there is no potential gain in binding ($\Delta=0$) or the monomers repel each other (negative $\Delta$), still in a sufficiently narrow tube they would prefer to bind. 

However, the equilibrium analysis leading to the Eq. \eqref{DF} cannot predict which is the length of bonded segment: obviously, in equilibrium, the larger the overlap $m$ is, the lower is the free energy. The reason why the two chains adopt a particular binding length $m^*$ is entirely kinetic. Two effects compete: the effective rate of the bonding reaction, and the rate of chain reptation along the tube to increase the overlap length $m$. Once any one of the $m$ bonds along the overlapped segment is established, the reptation stops, and the rest of the bonds set rapidly, in sequence, via the process of accelerated zipper \cite{Bell2017}. 

Let us say that the rate of an individual bonding reaction between two monomers is $k_\mathrm{on}$, determined by the chemistry involved. The probability of setting an individual bond during a given time interval $\Delta t$  is \cite{Eyring}: $
p_\mathrm{on} = 1-e^{-k_\mathrm{on}\Delta t}  $ 
(at small $\Delta t$, $p_\mathrm{on} =k_\mathrm{on}\Delta t$, while at large $\Delta t$, the reaction occurs with an almost certainty). The probability to establish at least one bond out of a possible $m$ sequence is
\begin{equation}
P(m) = 1-[1-p_\mathrm{on}]^m  = 1-e^{-m k_\mathrm{on}\Delta t(m)} .   \label{Pm}
\end{equation}
Any one of the $m$ possible bonds along the overlapping segment stops the reptation diffusion, and is rapidly followed by the full bonding of the segment. 

\begin{figure}
\includegraphics[width=0.95\columnwidth]{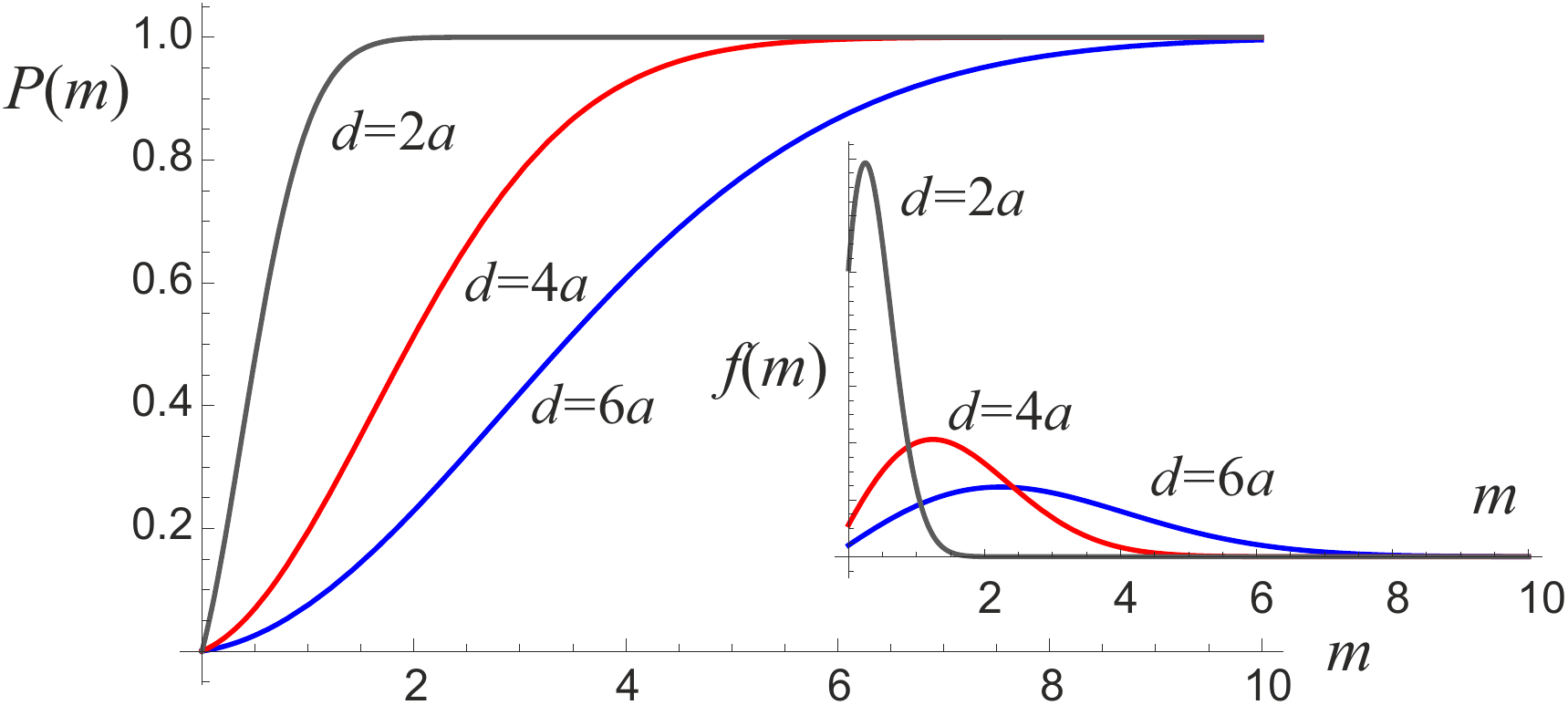}
\caption{\label{fig2} Plots of the cumulative probability $P(m)$, for the non-dimensional ratio $a^2k_\mathrm{on} /D_0 =10$, and several values of channel diameter, to illustrate that as the channel becomes more narrow and the chain reptation -- very slow, the two chains are likely to bind over very few units, just at the end. At larger $d$ the chains have time to move deeper into overlap before the binding occurs at a larger $m$. Inset: Corresponding plots of the probability density $f(m)$, Eq. \eqref{fm}.}
\end{figure}

We are now left with the second aspect of this kinetic problem: to find the time interval $\Delta t(m)$ for which the sequence of $m$ monomers dwells when two chains slide reptate past each other. The theory of polymer reptation is classical \cite{DoiEdwards}, and in our context we need to recall the characteristic time to diffuse the chain of $N$ monomers over a distance $ma$ in a tube of the remaining clearance $(d-a)$, cf. Fig. \ref{fig1}(b):
\begin{equation}
\tau[m] = \frac{(ma)^2}{D_c} , \ \ \ \  \mathrm{with} \ \ D_c=\frac{\pi^2k_BT}{N \gamma} \left( \frac{d-a}{a}\right)^2 , \label{tauM}
\end{equation}
where the diffusion constant for the chain centre of mass is given by $D_c$, in which $\gamma$ is the friction constant of just one monomer (so $D_0=k_BT/N\gamma$ is the diffusion constant of the whole polymer in free space: a value frequently measured for globular proteins in water). Equation \eqref{tauM} is more familiar in the form of reptation time of the whole chain length ($Na$) in a clear tube ($d$), when $\tau_d= N^3 a^4 \gamma/\pi^2 k_BT d^2$; see  \cite{DoiEdwards} for detail. We now have:
\begin{equation}
\Delta t(m) = \tau[m+1]-\tau[m] =\frac{N a^4 \gamma}{\pi^2k_BT (d-a)^2} (2m+1).  \label{tm}
\end{equation}
All these expressions from reptation theory are only valid for the tightly confined chain, with $d \ll R_g=N^{1/2}a$. 

\begin{figure}
\includegraphics[width=0.75\columnwidth]{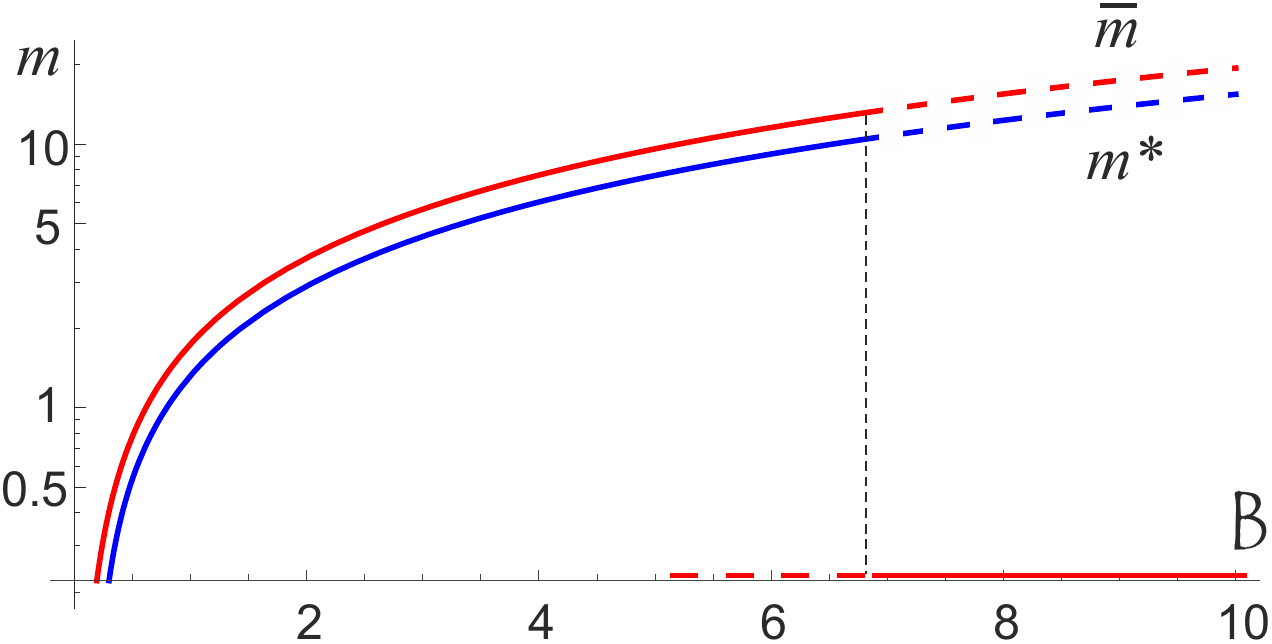}
\caption{\label{fig3}Comparison of the mean ($\bar{m}$) and the median ($m^*$) of $f(m)$, both expressions depending on the non-dimensional parameter ${\cal B}$ given in Eq. \eqref{mStar}. Note the logarithmic scale of the $m$-axis. The dashed line indicates the range of tube diameters $d$ above which the reptation theory is no longer valid, and the probability of $m$ monomers to overlap vanishes. }
\label{mfig}
\end{figure}

Figure \ref{fig2} illustrates how the probability $P(m)$ depends on its key parameters. Now the kinetic problem we are addressing becomes that of the mean first-passage time \cite{MFPT,MFPT3}. The question is: at which $m^*$ the first binding reaction would occur. The probability density that binding occurs at a given value of overlap $m$ is given by $f(m) = dP(m)/dm$, so that the mean first-binding length is $\bar{m} = \int m f(m) \mathrm{d}m$. Carrying out the algebra, we obtain $f(m)$ illustrated in Fig. \ref{fig2} inset:
\begin{equation}
f(m)= \frac{(4m+1)a^4 k_\mathrm{on}}{\pi^2 D_0(d-a)^2} \exp \left[- \frac{m(2m+1)a^4 k_\mathrm{on}}{\pi^2D_0(d-a)^2} \right].    \label{fm}
\end{equation}
The peak of this probability distribution is determined by a single non-dimensional parameter, which we may call `bonding enhancement' ${\cal B}$:
\begin{equation}
m^* = \frac{\pi}{2} {\cal B} -\frac{1}{4}, \ \ \ \mathrm{with} \ \  {\cal B} =\frac{d-a}{a} \sqrt{\frac{D_0}{a^2k_\mathrm{on}}}.  \label{mStar}
\end{equation}
The mean bonded length $\bar{m}$ is also easily calculated, but its expression is more cumbersome. Figure \ref{fig3} shows that there is no great difference in values between the mean and the median of $f(m)$, and for all practical purposes we may work with the simpler expression in Eq. \eqref{mStar}.

It is interesting to assess the critical diameter of the tube, $d^*$, below which the bonding is likely. It is obtained by solving the transcendent equation $\Delta F(d)=0$, and has the form
\begin{equation}
\frac{d^*}{a} = \sqrt{ \frac{\pi^2 m^*/3}{\mathrm{ProductLog}[\pi^2 m^*e^{-m^*\beta \Delta}/3]}} \approx e^{m^* \beta \Delta/2},
\label{dS}
\end{equation}
where $m^*$ is given by Eq. \eqref{mStar}, and 
the approximate form is valid when $m^*e^{-m^*\beta \Delta} \ll 1$ (i.e. essentially for all non-negative $\Delta$). The exponential approximation of Eq. \eqref{dS} is prominent for all non-repulsive interactions, while for negative $\Delta$ (repulsion between monomers) the other limit of `ProductLog' is in force, and $d^* \sim a$ (in other words, bonding is unlikely). However, there is another upper boundary for the channel diameter that promotes binding: all analysis in Eqs. (\ref{tm}-\ref{fm}) was based on the chain reptation dynamics, which stops being applicable at $d \geq N^{1/2}a$ when the random-coil (blob) chain dynamics would take over (the exact crossover is hard to identify). This is schematically labelled in Fig. \ref{mfig}.

Assuming we are in narrow enough channels for the reptation theory to work, we should try estimating the kinetically-set bonding length $m^*$, which requires the value of parameter ${\cal B}$. The friction constant for one monomer (e.g. aminoacid) can be obtained approximately from $\gamma = 6\pi \eta a$, with
$\eta$ being the viscosity of water (0.7\,mPa.s), and $a=0.3$\,nm the size of a residue \cite{binding}. A flagellin protein with $N \approx 495$, at room temperature, should have the diffusion constant:
{$D_0 = 2 \cdot 10^{-12} \mathrm{m}^2/$s (in agreement with a typical diffusion constant of aminoacids in water \cite{diffu1,diffu2}: $k_BT/\gamma \sim 10^{-9} \mathrm{m}^2/$s, divided by 500). }

It is hard to find values of the bonding reaction rate $k_\mathrm{on}$ for a pair of aminoacid residues: it strongly depends on a large number of specific chemical factors \cite{rates}. In the spirit of our `average polymer chain' with a single characteristic value of bond energy gain $\Delta$, we may try a basic Kramers estimate: $k_\mathrm{on} = \omega_0 e^{\beta \Delta}$, where $\omega_0$ is the collision frequency in solution: $\omega_0 \approx 10^{8} \, \mathrm{s}^{-1}$ \cite{Eyring,Laidler}. Then, for a low $\beta \Delta =1$, we have $k_\mathrm{on} = 2.7 \cdot 10^{8} \, \mathrm{s}^{-1}$. Using Eq. (\ref{mStar}) with the diameter of flagellar channel $d=2$\,nm, we obtain a value: ${\cal B}=1.63$, and the preferred length of binding $m^* = 2.3$. As a consequence, for the same set of parameters, the free energy of binding is very low: $\Delta F \approx -k_BT$.  A high reaction rate would commit the constrained chains to bond on the first monomer ($m^*\leq 1$), and the bonding free energy is determined purely by the potential energy of binding $\beta \Delta$.  For comparison, a lower reaction rate $k_\mathrm{on} = 10^{7} \, \mathrm{s}^{-1}$ would give: ${\cal B}=8.4$ and $m^* = 13$, with the coresponding binding free energy $\Delta F \approx -12k_BT$ (maintaining  $\beta \Delta =1$). 

In order to test the ideas of the confinement-induced bonding, we carried out simulation of two chains in a reflective  channel, using the LAMMPS package. We used the Kremer-Grest model for coarse-grained polymers \cite{kremer_grest,everaers} where bonds between consecutive monomers are modelled with a harmonic-like finitely extensible elastic potential, made of a combination of the attractive FENE part and the repulsive LJ part (the approach used in many studies, old and new \cite{ba1,ba2}), and the weak bending potential with the constant $k_\theta= k_BT$. The non-bonded particles interact through the LJ potential of variable weak attraction strength (see \cite{Lappala,Lappala2}). Two kinds of LJ potential create a subtlety in the simulation setup: one needs to be clear which `LJ units' are used. We kept  the average temperature constant, equal to the attractive strength $\varepsilon$ of the pair LJ potential (which is equivalent to our earlier condition $\beta \Delta =1$). The LJ time and the damping constant of the simulation were linked to the same energy scale via the fluctuation-dissipation theorem: $\tau = \sigma \sqrt{M/\varepsilon}$ and \textit{damp}$=1/\tau$, where $M$ is the mass and $\sigma$ the diameter of an individual monomer, equivalent to the parameter $a$ used in the analysis above (the simulation time-step was $dt=0.01\tau$). On the other hand, in the Kremer and Grest model, the bond potential has the FENE bond strength linked to the repulsive-LJ strength: $k=30\varepsilon^*/\sigma^2$ \cite{kremer_grest,everaers}, which makes the bond length $r_\mathrm{bond} \sim 0.96\sigma$. 

\begin{figure}
\includegraphics[width=0.75\columnwidth]{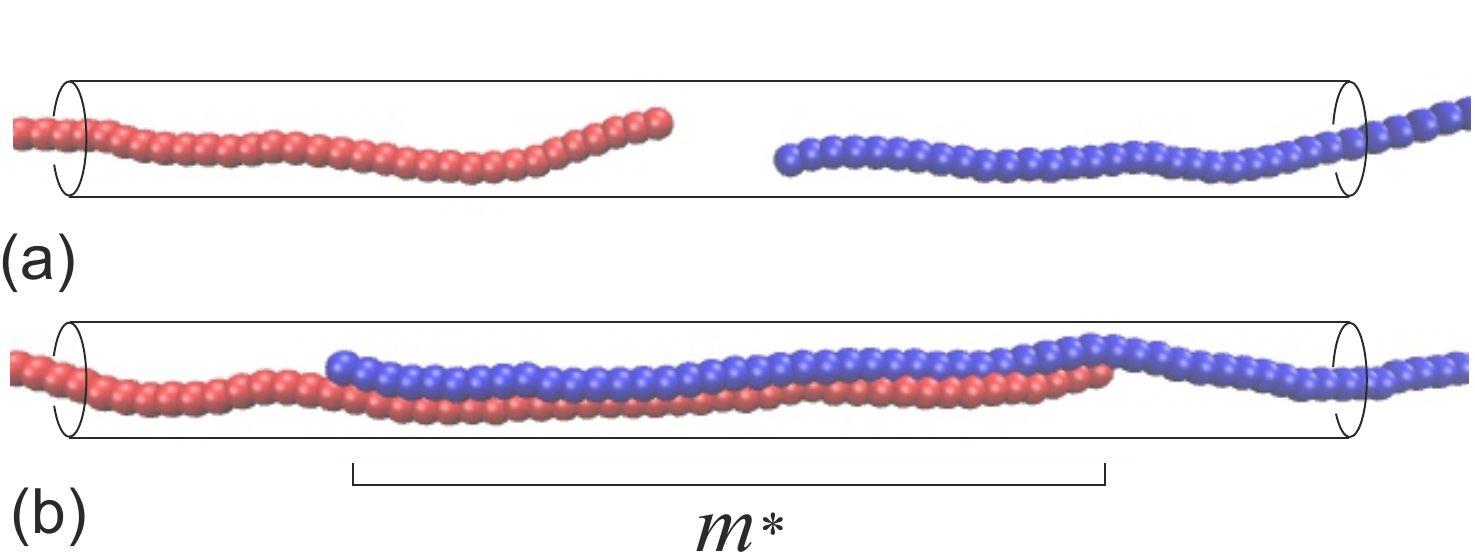}
\caption{\label{tubes} Simulation snapshots of two chains in the channel  $d=3\sigma$.  All simulations started with chains separated by $6\sigma$; the overlap number $m^*$ at different times is shown in Fig. \ref{mSfig}.}
\end{figure}

\begin{figure}
\includegraphics[width=0.85\columnwidth]{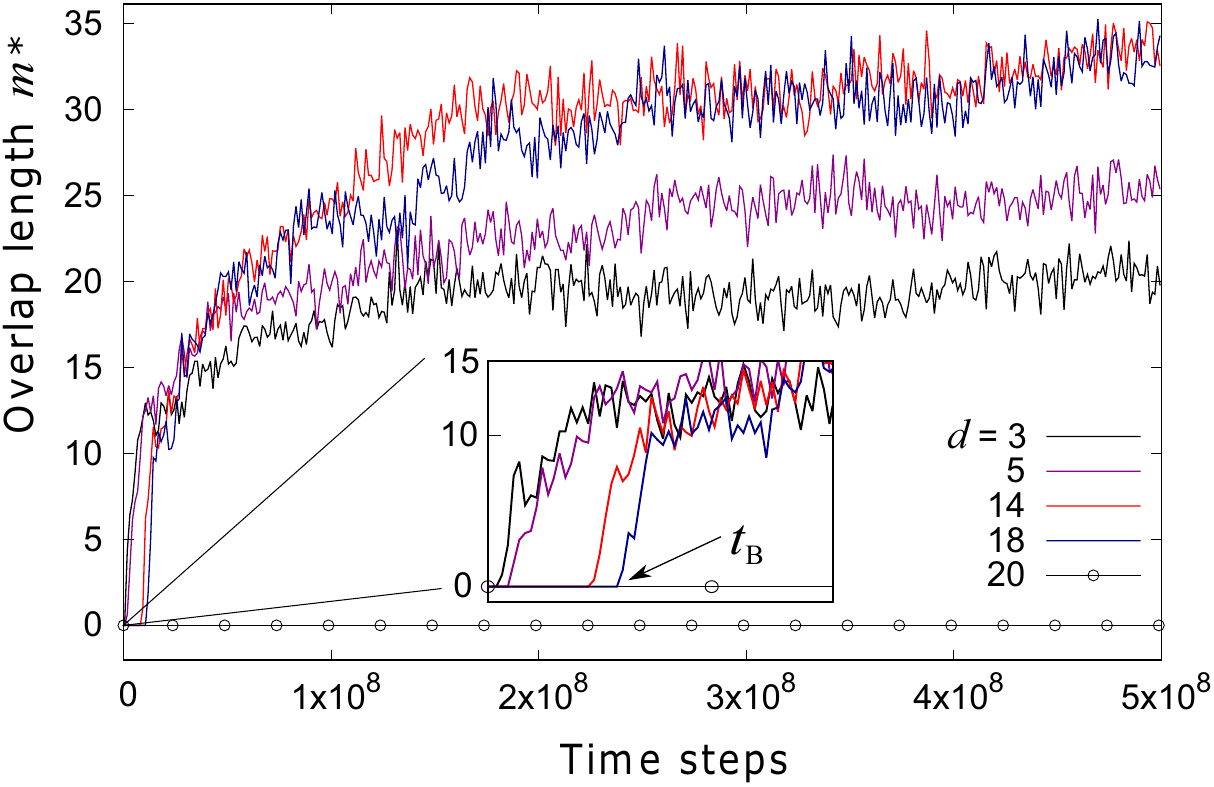}
\caption{\label{mSfig} The overlap distance $m^*$ of two long chains in a tube with different values of $d$. We determined $m^*$ as the number of monomers, which were $r \leq 1.1225\sigma$ apart: just above the minimum of the LJ potential at $2^{1/6}\sigma$. The time $t_\mathrm{B}$ when the two chains first `lock' is shown in the inset. The most frequent binding time $\overline{ t_\mathrm{B} }$ increases rapidly when the tube diameter increases.} 
\end{figure}

However, we found that when the LJ repulsion strength $\varepsilon^*=\varepsilon$ (which, in turn, is equal to $k_BT)$, the chains do not bind in the channel. The reason was that the length of the bond between monomers on the chain was fluctuating strongly, and the required `matching' of monomers along the two chains was not possible. When we insisted that the bond along the chain is strongly confined, by taking $\varepsilon^*=100 \varepsilon$, the confinement-induced binding of two chains was a strong and obvious effect, see Fig. \ref{tubes}. We observed that tightly confined chains eventually come in contact during their reptation diffusion; mostly such a contact was short-lived and the chains separated again -- but on rare occasions, when a sequence of several monomers come into contact simultaneously, the chains `lock' in binding. After a long simulation time, the number of overlapping monomers $m^*$ was reaching the equilibrium value, depending on the channel diameter, see Fig. \ref{mSfig}. Note that in these simulations the $m^*$ plateau value for $d=14a$ and $18a$ was the same; this was a systematic result reproduced in a large number of independent simulation runs, for all large diameters where the binding occurred. For $d=20a$ we did not register any binding event within the given time of simulation.

\begin{figure}
\includegraphics[width=0.75\columnwidth]{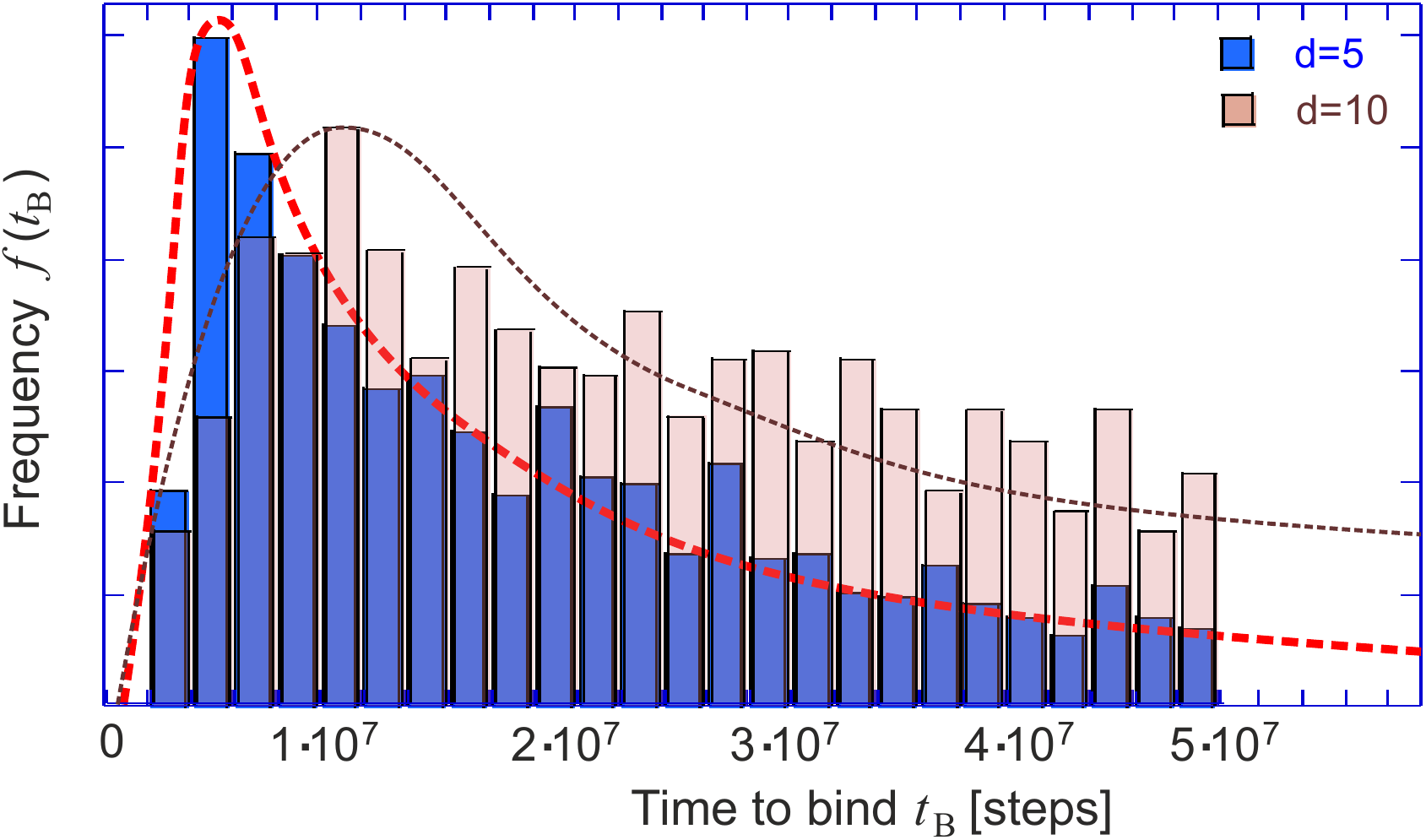}
\caption{\label{histo} Distribution of binding times $f(t_\mathrm{B})$ for $\beta \Delta=1$. The main dataset is for $d=5a$, with the incomplete data for $d=10a$ in the background. Statistics is on $2000$ independent simulations of $5\cdot 10^7$ time steps. Even in the narrow channel, we are far from capturing all the rare events in the simulation time; at $d=10a$ only about a half of all simulations ended in binding. Dashed lines are the model distribution, with the decaying tail fitted to $t^{-0.8}$ power law. } 
\end{figure}

As the stimulated-binding process is clearly driven by chain kinetics in the confining tube, we have done many simulations (with different random seeds) to establish the statistical distribution of binding time $t_\mathrm{B}$. Figure \ref{histo} gives an illustration of such a distribution, $f(t_\mathrm{B})$, obtained for $d=5a$; the plot also gives the result for $d=10a$ in the background, but we were not able to acquire a sufficient statistics for larger $d$ because of too many instances of very long $t_\mathrm{B}$. This distribution has a very characteristic `heavy tail': it decays with a weak power law (our fitting suggests $t^{-0.8}$). One of the consequences is that the average binding time $\langle t_\mathrm{B} \rangle = \int  t_\mathrm{B} f( t_\mathrm{B}) d  t_\mathrm{B}$ is not defined (the integral diverges), and we have to use the median $\overline{ t_\mathrm{B}}$ instead. Such distributions are found in the Levy flight processes \cite{Levy1,Levy2}, and in our case the origin is similar: the events leading to an overlap of $m^*$ monomers are rare, but only these lead to the lasting binding of the chains. The key point is clear: two chains with very weak attraction (i.e. not naturally prone to binding on contact in the bulk) would be strongly bound if a large number of their monomers would simultaneously overlap. In a weakly constrained case, the probability of such overlap is very low, and the time $t_\mathrm{B}$ to wait for such a rare event is longer than any realistic experimental time.

The theory presented here is deliberately qualitative, demonstrating the key concept of stimulated bonding in narrow channels. Its analysis is essentially scaling, with explicit calculations based on the ideal Gaussian chain limit (however, this limit becomes accurate for polymer chains under strong lateral confinement, such as in dense melts or other `tube-model' situations). The idea of an `average' homopolymer chain is also very limiting: a real protein would have a variety of aminoacid residues with very different potential energy of pair binding $\Delta$, between $-4k_BT$ and $+4k_bT$ (see \cite{binding,binding2} for detail): a follow-up problem with the broadly distributed quenched $\Delta$ is certainly of interest. Nevertheless, these approximations and simplifications allowed us to expose the physics of entropically stimulated bonding, and obtain analytical expressions for the main points of interest. 

It is possible that a similar analysis would predict a much enhanced protein folding, once the conformational entropy is reduced by the tight channel confinement, as in ribosomes \cite{ribosome1,ribosome2}. Another interesting corollary of this concept is the need to examine effective reaction rates in micro-volumes. Today the technology of microfluidics and a `lab on a chip' is widespread: the reduction of translational entropy (and when unfolded polymer chains are involved, also the configurational entropy) needs to be taken into account when analysing the reaction rates in very small volumes.  \\

\begin{acknowledgments}
We thank Samuel Bell and Cheng-Tai Lee for very helpful discussions. GMF acknowledges support from the BBSRC (BB/M007197/1). MSGF acknowledges support from 
Coordena\c c\~ao de Aperfei\c coamento de Pessoal de N\'ivel Superior, Brasil (CAPES), Fin. Code 001. EMT acknowledges support from the ERC AdG 786659 (APRA). 
\end{acknowledgments}

\end{document}